# Real-time superresolution interferometric measurement enabled by structured nonlinear optics

Xin-Yu Zhang[1,2], Hai-Jun Wu[1], Bing-Shi Yu[1], Carmelo Rosales-Guzmán[1,3], Zhi-Han Zhu[1*], Xiao-Peng Hu[2*], Bao-Sen Shi[1,4*] and Shi-Ning Zhu[2]

[1] Wang Da-Heng Center, HLJ Key Laboratory of Quantum Control, Harbin University of Science and Technology, Harbin 150080, China
[2] National Laboratory of Solid-State Microstructures, College of Engineering and Applied Sciences, Nanjing University, Nanjing 210093, China
[3] Centro de Investigaciones en Óptica, A.C., Loma del Bosque 115, Colonia Lomas del Campestre, 37150 León, Gto., Mexico
[4] CAS Key Laboratory of Quantum Information, University of Science and Technology of China, Hefei, 230026, China
* e-mail: zhuzhihan@hrbust.edu.cn, xphu@nju.edu.cn, and drshi@ustc.edu.cn

**Abstract**: Optical interferometers are pillars of modern precision metrology, but their resolution is limited by the wavelength of the light source, which cannot be infinitely reduced. Magically, this limitation can be circumvented by using an entangled multiphoton source because interference produced by an $N$-photon amplitude features a reduced de Broglie wavelength $\lambda/N$. However, the extremely low efficiency in multiphoton state generation and coincidence counts actually negates the potential of using multiphoton states in practical measurements. Here, we demonstrate a novel interferometric technique based on structured nonlinear optics, i.e., parametric upconversion of a structured beam, capable of superresolution measurement in real time. The main principle relies in that the orbital angular momentum (OAM) state and associated intramodal phase within the structured beam are both continuously multiplied in cascading upconversion to mimic the superresolved phase evolution of a multiphoton amplitude. Owing to the use of bright sensing beams and OAM mode projection, up to a 12-photon de Broglie wavelength with almost perfect visibility is observed in real time and, importantly, by using only a low-cost detector. Our results open the door to real-time superresolution interferometric metrology and provide a promising way toward multiphoton superiority in practical applications.

## Introduction

In 1887, after six years of effort, Albert Michelson and Edward Morley announced their failure to observe the ether. This failure, however, provided the bases for Einstein to develop his special theory of relativity and began an era of optical interferometric measurement [1]. Since then, the optical interferometer has gradually become the primary toolkit of modern precision measurement. With the birth and evolution of laser techniques [2], significant progress has been made in increasing the precision of interferometers over the past half-century. By estimating the phase displacement of a laser interferometer, one can measure a great variety of quantities, such as position, speed, acceleration, spectrum, and medium properties, with ultrahigh precision [3–6]. In particular, the two largest ever Michelson interferometers were constructed to search for ripples in space-time as the core part of the Laser Interferometer Gravitational Wave Observatory [7]. This giant instrument has unprecedented precision and successfully observed gravitational waves for the first time in 2015, 130 years after the Michelson–Morley experiment, finally confirming the space-time view in Einstein's general theory of relativity. Despite hundreds of kilowatt lasers circulating within it, the performance of this giant interferometer has surprisingly already been limited by quantum theory [8–11].

The performance of any interferometer depends heavily on its phase resolution and sensitivity, which are limited by the de Broglie wavelength and shot-noise limit of the light source, respectively [12,13]. Interestingly, both these limitations can be overcome by using $N$-photon entangled states as the light source, allowing $N$-fold superresolution and sensitivity in interferometric measurement. The underlying principle is, first, that the de Broglie wavelength of an $N$-photon wave packet, such as the reputed $N$00$N$ state, depends on both the wavelength ($\lambda$) and number ($N$) of photons, i.e., $\lambda/N$. Second, the shot-noise limit originates from the uncertainty relation between the amplitude (i.e., number of photons) and phase of the light field, i.e., $\Delta\varphi\Delta n \geq 1$. In $N$00$N$ states, the largest uncertainty is in the number of photons, corresponding to a minimal phase uncertainty, and these states can therefore be used to approach the Heisenberg limit in phase estimation, given by $\Delta\varphi = 1/n$ [14]. Notably, pursuing supersensitivity in an infinitesimal area near the balance phase position, i.e., $cos(\pi/2)$, of a SU(2) interferometer is only meaningful if perfect transmission-detection efficiency and sufficient light power (such as kilowatt lasers) are attainable [3,10,15]. Therefore, squeezed light (having no superresolution effect), not an $N00N$ state, was chosen to further enhance the sensitivity of the detecting gravitational-wave detector [16,17]. In contrast, realization of superresolution interferometric measurement featuring a reduced de Broglie wavelength with low-cost detectors and especially in a real-time manner would be beneficial in various applications [18,19].

Measuring a reduced de Broglie wavelength attracts constant interest in the quantum optics community [20–22], ranging from the most often demonstrated two-photon interference to the later reported 3-, 4- and 8-photon experiments and the recently observed 10- and 12-photon entanglement [23–29]. These demonstrations, on the one hand, are important benchmarks that show the considerable progress made in multiphoton manipulation, which is the physical basis for building photonic quantum information systems [30–31]. On the other hand, the extremely low multiphoton coincidence rate (several photons per hour) and the poor interference visibility shown in results negate the potential advantage of multiphoton states in interferometric measurement. To date, only one recent two-photon experiment has realized an acceptable coincidence rate and high visibility by using expensive superconducting detectors and photon pairs at a specific wavelength (coherence length matched) [32]. To avoid using the inefficiently multiphoton state, scientists have also tried to directly extract multiphoton amplitude from a classical sensing beam [33]. But the efficiency of post-selecting multiphoton states from weak cohere light is still too low, as a result, an extremely low coincidence rate is still an insoluble problem. Overall, the realization of superresolution superiority with both high signal power and high visibility remains a challenge. The bottleneck originates from the intrinsic inefficiency in multiphoton state generation, manipulation and detection, which has not yet been overcome and is also impossible to break in the short term.

In this work, we present a novel interferometric technique capable of superresolution phase measurements in real time. In contrast to the low efficiency in generating and detecting multiphoton states, our work uses an efficient approach to build and measure a multiphoton amplitude, i.e., parametric upconversion of structured light carrying phase information ($\varphi$) to be measured. Structured light refers to a spin-orbit coupling (SOC) state, and $\varphi$ is coded in its intramodal phase by a birefringence shifter. Crucially, the orbital angular momentum (OAM) and associated intramodal phase ($\varphi$) of the SOC state are both multiplied in the upconversion. By using cascading parametric conversion, as well as OAM mode post-selection, we measure, in real time, up to $N = 12$ times narrower superresolution fringes ($\lambda/N$) with almost unit visibility. In the following, we experimentally demonstrate this technique scheme.

## Results

**Concept and Principle** — The structured light involved in this work refers to a cylindrical vector (CV) beam with spatially varying polarization [34]. The beam constitutes a nonseparable superposition state of orthogonal circular polarizations $\hat{e}_\pm$ and spatial modes $\psi_{\pm\ell}$ carrying opposite OAMs $\pm\ell\hbar$; the most common Laguerre–Gaussian modes are considered here [35]. This paraxial SOC state can be simply represented by using the Dirac notation [36]:

$$|\Psi_{soc}\rangle = \sqrt{\frac{1}{2}}(|\hat{e}_+, \psi_{+\ell}\rangle + e^{i\varphi}|\hat{e}_-, \psi_{-\ell}\rangle), \qquad (1)$$

where $\varphi$ is the intramodal phase between the two polarization components. The mathematical form of the SOC state is similar to that of the path-number (i.e., $N00N$) state in an SU(2) interferometer. Thus, this classical "entangled" state has been widely used to mimic and study the quantum behavior of entanglement [37–42]. In addition, the spatial polarization structure of state (1) is $2\pi/\ell$ periodic in the azimuthal direction, giving rise to an $\ell$-fold angular sensitivity in an image rotation operation. For this reason, state (1) has an $N00N$-like superiority in rotation angle measurements, i.e., the phase variation becomes $\ell\Delta\varphi$, but this principle is only valid in the scenario of image rotation [43].

Therefore, an important question is how can the $N00N$-like superiority of this "entangled" bright beam be realized in more general interferometric measurement? Namely, without exploiting its $\ell$-fold angular sensitivity, how can the behavior of the $N$-photon amplitude and associated phase variation $N\Delta\varphi$ be built and recorded? We propose cascading second-harmonic generation (SHG) of the sensing structured beam to continuously double both the OAM and associated intramodal phase, mimicking an $N00N$-state injected interferometer. Specifically, we represent the SOC state using orthogonal linear polarizations $|\hat{e}_{h,v}\rangle = \sqrt{1/2}\,(|\hat{e}_+\rangle \pm |\hat{e}_-\rangle)$, i.e., the mutually unbiased basis of circular polarizations, so that state (1)

becomes $\sqrt{1/4}\,[(|\psi_{+\ell}\rangle + e^{i\varphi}|\psi_{-\ell}\rangle)|\hat{e}_h\rangle + (|\psi_{+\ell}\rangle - e^{i\varphi}|\psi_{-\ell}\rangle)|\hat{e}_v\rangle]$. We then perform type-II SHG for this state, or rather, sum-frequency generation (SFG) between the state (1) $\hat{e}_h$- and $\hat{e}_v$-components [44], leading to, in addition to frequency doubling ($2\omega$), a 2nd-harmonic OAM mode:

$$|\Psi_{N=2}\rangle = \sqrt{\frac{1}{2}}(|\psi_{+2\ell}\rangle - e^{i2\varphi}|\psi_{-2\ell}\rangle). \qquad (2)$$

Here, the $2\varphi$ phase-evolution behavior within the modal space spanned by conjugate OAM modes $\psi_{\pm 2\ell}$ is exactly the same as that in an SU(2) interferometer injected by the $N00N$ state with $N = 2$. On this basis, we can further achieve three- or fourfold superresolution by performing SFG between the state (2) and state (1) $\hat{e}_v$-components or SHG of state (2), respectively. The corresponding harmonic OAM modes are:

$$|\Psi_{N=3}\rangle = \sqrt{\frac{1}{4}}(|\psi_{+3\ell}\rangle + e^{i3\varphi}|\psi_{-3\ell}\rangle)$$
$$- \sqrt{\frac{1}{4}}(e^{i\varphi}|\psi_{+\ell}\rangle + e^{i2\varphi}|\psi_{-\ell}\rangle), \qquad (3)$$

$$|\Psi_{N=4}\rangle = \sqrt{\frac{1}{4}}(|\psi_{+4\ell}\rangle + e^{i4\varphi}|\psi_{-4\ell}\rangle) - \sqrt{\frac{1}{2}}e^{i2\varphi}|\psi_0\rangle \qquad (4)$$

The $3\varphi$ and $4\varphi$ phase-evolution behaviors within the 3rd- and 4th-harmonic OAM modes (i.e., $\psi_{\pm 3\ell}$ and $\psi_{\pm 4\ell}$) provide an interface to observe the superresolved interference resulting from the three- and four-photon de Broglie wavelengths of the fundamental waves, respectively. Compared with the superresolved interference obtained from high-$N00N$ states, the difference here, as well as its core advantage, is the use of a bright structured beam as the sensing channel. Consequently, there is no need for expensive and inefficient photon counters for multiphoton coincidence. In addition, similar to the issue encountered when using a high $N00N$ state with $N > 2$ [25], unwanted OAM modes, such as $\psi_{\pm \ell}$ and $\psi_0$ contained in states (3) and (4), become noise and remain in the upconverted waves. However, by using OAM mode projection [45], we can easily extract the interference between OAM modes of interest (i.e., $\psi_{\pm 3}$ and $\psi_{\pm 4}$) and thus achieve near-perfect interference visibility.

Note that we cannot attribute the superresolution obtained here to a reduction in the wavelength of upconverted waves despite this occurring. In contrast, shorter wavelengths make the signal hard to control, hindering the pursuit of higher-ratio superresolution. Luckily, this technique framework allows us to approach this problem straightforwardly, i.e., by exploiting parametric downconversion to roll back the wavelength of the signal, which will be demonstrated later with specific experiments in the paper.

**Experimental Results** — To test the above principle, we performed a series of proof-of-principle experiments. Figure 1a shows the schematic setup of the superresolution interferometric measurement for $N = 2, 3,$ and $4$. By using a q-plate combined with a half-wave plate (birefringent phase shifter), a horizontally polarized Gaussian beam with a wavelength of 1560 nm was converted into state (1) with $\ell = 1$ and adjustable intramodal phase $\varphi$. The prepared SOC state was first characterized using a spatial Stokes polarimeter to determine $\varphi$ [23] and was then focused into three different quasi-phase-matching crystals to generate states (2), (3), and (4). Among the crystals, only that for the $N = 2$ experiment has a monoperiodic structure designed for solo type-II SHG (1560 nm $\hat{e}_h$ + 1560 nm $\hat{e}_v \rightarrow$ 780 nm $\hat{e}_h$), while the other two (for the $N = 3$ and $4$ experiments) have quasiperiodic structures designed to be compatible with dual upconversion [46–48], i.e., type-II SHG cascading type-0 SFG (1560 nm $\hat{e}_h$ + 780 nm $\hat{e}_h \rightarrow$ 520 nm $\hat{e}_h$) or type-0 SHG (780 nm $\hat{e}_h \times 2 \rightarrow$ 390 nm $\hat{e}_h$). The generated 2nd-, 3rd-, and 4th-harmonic waves were measured by OAM mode projection using spatial light modulation [49]. As a result, the signal from the superresolved interference within the $N$-harmonic OAM modes, i.e., $\sqrt{1/2}(|\psi_{+N}\rangle + e^{iN\varphi}|\psi_{-N}\rangle)$, appears at the center of far-field patterns of beams measured by spatial light modulator, which were recorded by a camera. More details on the experimental setup, nonlinear crystal parameters, and spatial mode transformation in the nonlinear interactions and projective measurements are introduced in the *Methods* and *Supplementary Materials*.

Figure 1b shows recorded interferometric signals, confirming the superresolution behavior of the intramodal phase within the $N$-harmonic OAM modes. The evolution of $e^{iN\varphi}$ ($N = 2, 3,$ and $4$), extracted from states (2), (3), and (4), compared with that in the original signal state (1) shown in the first row oscillates two-, three- and fourfold faster, respectively. All the results exhibit near-perfect interference visibility ($v \approx 1$), and more importantly, they are bright signals ($\sim mW$) recorded in real time using only a low-cost detector.

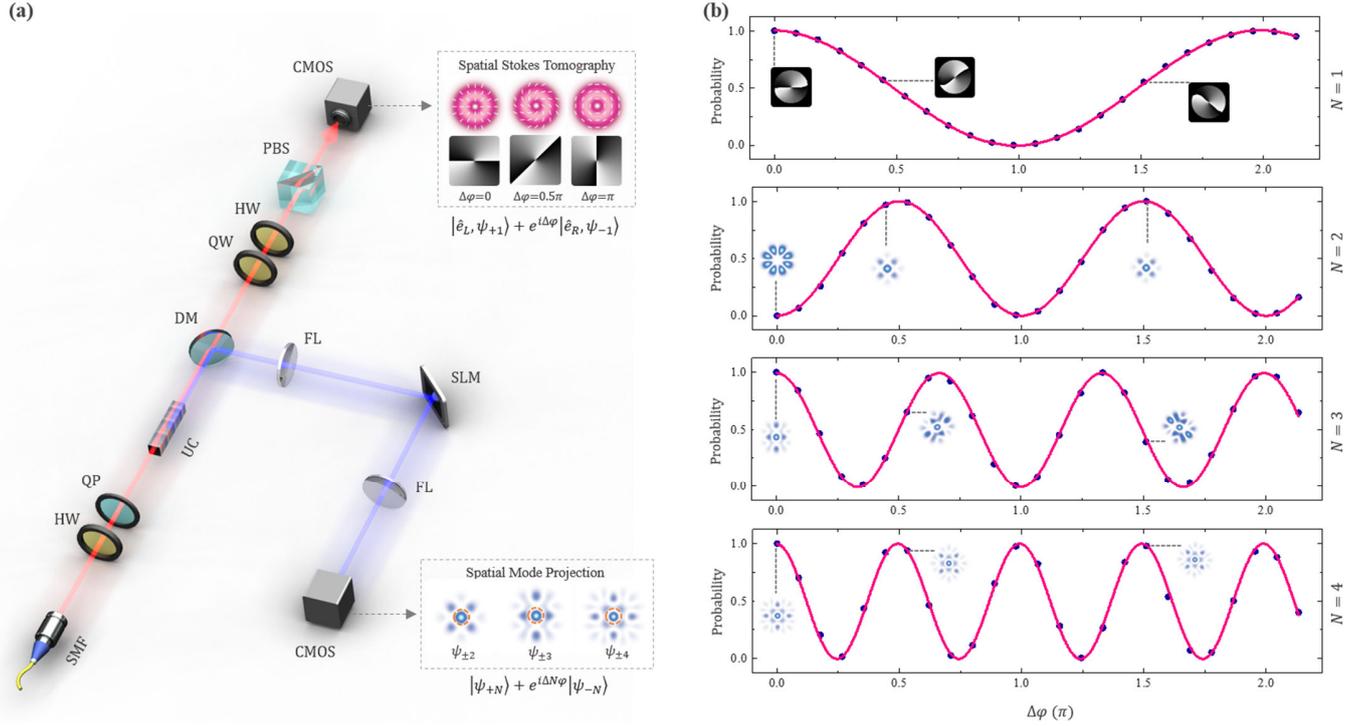

**Figure 1.** Experiments of superresolution interferometric measurements for $N$ = 2, 3, and 4. **(a)** Schematic of the experimental setup, where the key components are the single-mode fiber collimator (SC), half-wave plate (HW), quarter-wave plate (QW), q-plate (QP), upconversion crystal (UC), polarizing beam splitter (PBS), dichroic mirror (DM), Fourier lens (FL), spatial light modulator (SLM), and camera (CMOS). The original phase variation $e^{i\Delta\varphi}$ within the SOC state was determined via spatial Stokes tomography, as shown in the top inset, where samples in the first and second rows are vector profiles and associated spatial intramodal phases, respectively. The superresolved phase variation $e^{i\Delta N\varphi}$ was measured via spatial mode projection, as shown in the bottom inset, where the center patterns surrounded by orange dashed circles are the amplitude of interference. **(b)** Measured superresolved interference between conjugate OAM modes $\psi_{\pm N}$ with $N$ = 2, 3, and 4.

As mentioned above, owing to the diminishing wavelength of upconverted waves, the ratio $N$ cannot be further improved by directly cascading upconversion more times. For instance, the wavelength of the signal carrying $e^{i4\varphi}$ obtained in the above experiment is 390 nm, which has already arrived at the edge of the ultraviolet spectrum. Therefore, using parametric downconversion to roll back the wavelength of upconverted waves is a straightforward but effective approach. Figure 2a shows the schematic setup of the approach (see *Supplementary Materials* for more details). We performed type-II degenerate downconversion for the signal $N = 4$ at 390 nm to roll back its wavelength to 780 nm and then cascaded type-0 SHG to obtain a superresolved signal with a ratio $N = 8$. Similarly, for the signal $N = 3$ at 520 nm, we used nondegenerate downconversion to roll back its wavelength to 1560 nm and then cascaded SHG another two times, $1560\ nm \rightarrow 780\ nm$ and $780\ nm \rightarrow 390\ nm$, obtaining superresolved signals with ratios $N = 6$ and $N = 12$, respectively. In addition, to maintain the transverse structure of rolled back beams, a super Gaussian pump was used in the downconversion [50]. Details on involved spatial modes and associated mode transformation are given in the *Supplementary Materials*.

Figure 2b shows the observed interferometric behavior within the $N$-harmonic OAM modes $N = 6,\ 8,$ and 12, which oscillate six, eight, and even 12 times faster, respectively, than the original state (1); moreover, all the signals have near-perfect visibility. To clearly show the benefits of high-ratio superresolution, the phase variation range was chosen to be in the insensitive region of the original sensing signal, i.e., near $cos\pi$. Compared with the almost constant amplitude of the original signal, all the superresolved signals oscillate for more than one cycle, and all the signal powers are visible to the naked eye. With respect to the data shown in the last row, in particular, *this is the first observation of a 12-photon de Broglie wavelength in real time*. In comparison, observing a reduced de Broglie wavelength of entanglement composed of approximately 10 photons, even with state-of-the-art techniques, requires several hours to record each data point of the photon-coincidence interference and has poor visibility [26–29].

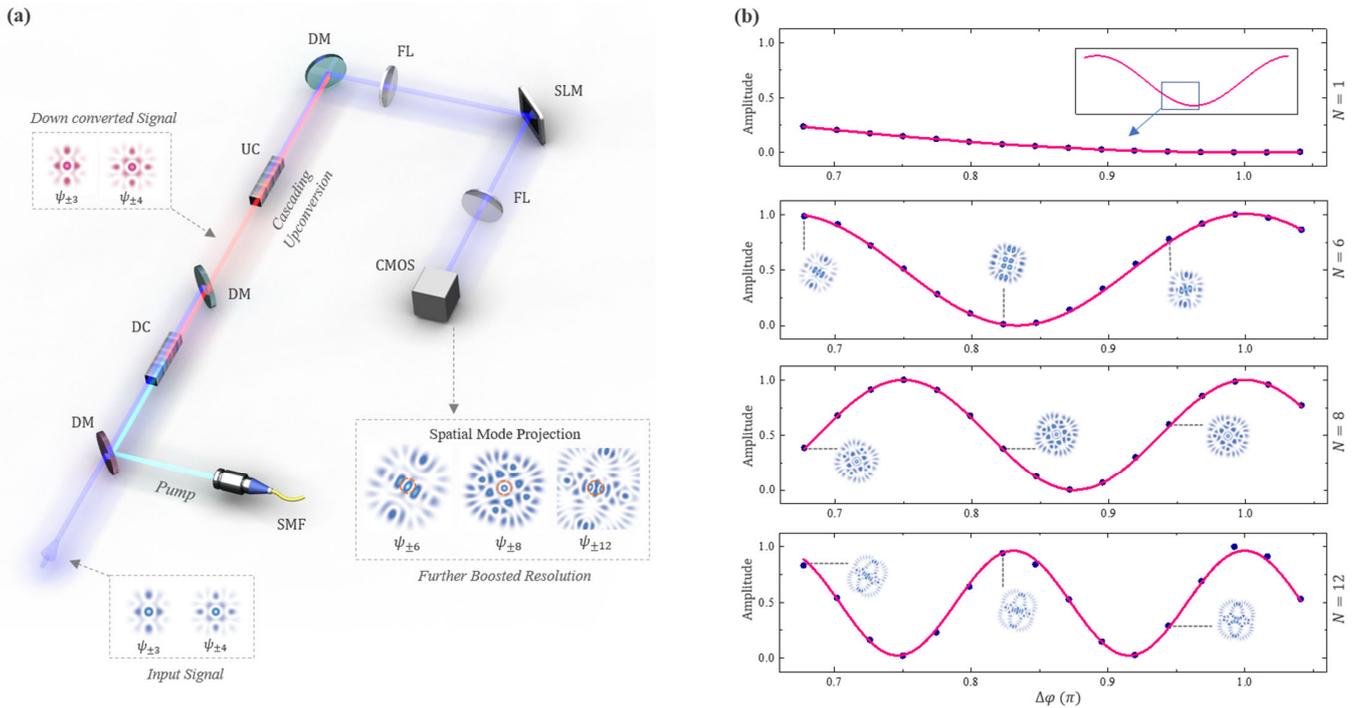

**Figure 2.** Experiments of superresolution interferometric measurements for *N* = 6, 8, and 12. **(a)** Schematic of the experimental setup, where DC denotes the crystal for downconversion, and the other components are the same as those in Fig. 1(a). **(b)** Measured superresolved interference between conjugate OAM modes $\psi_{\pm N}$ with *N* = 6, 8, and 12.

## Discussion

We have demonstrated a series of superresolution interferometric measurements using nonlinear interactions of structured light. Up to a dozen-fold superresolved interference, corresponding to a reduced de Broglie wavelength of $\lambda/12$, was successfully observed in real time and with near-perfect visibility. This scheme provides a novel roadmap for the development of advanced interferometric techniques with ultrahigh phase resolution. Compared with previous work [23–29,32,33], our results show two impressive advances: real-time measurement (high-power signal) and near-perfect visibility ($v \approx 1$), enabled by the use of bright sensing beams and OAM mode projection, respectively. Notably, the latter gives rise to a natural question: does the phase sensitivity in this scheme also exceed the shot-noise limit? If we only consider the threshold criteria of the shot-noise limit with respect to interference visibility, then the answer seems yes. However, this conclusion will be negated when the energy loss of the final signal relative to the original sensing beam passing through the phase shifter is considered, as was true for almost all previous superresolution work [25]. Regarding the energy loss in this scheme, first, the weight of $e^{iN\varphi}$-carrying modes (i.e., $\psi_{\pm N\ell}$) in the upconverted wave becomes less than one when $N > 2$ and continuously decreases with *N*; second, the upper bound of the SHG efficiency in theory is also less than one (*see Supplementary Materials*). Therefore, the detectable signal power of $\psi_{\pm N\ell}$ extracted from the final upconverted wave is much lower than that of $\psi_{\pm\ell}$ in the original sensing beam, especially for a high-*N* case.

Despite this, the absolute phase sensitivity depending on detectable signal energy obtained here is still far greater than that in any previous work involved superresolution. To date, only one recent study using state-of-the-art superconducting nanowire single-photon detectors has demonstrated an *N*00*N*-state interferometer that can unconditionally surpass the shot-noise limit (i.e., the evaluation included transmission-detection loss) [32]. However, only a superresolution ratio of *N* = 2 was achieved in that study, equivalent to a de Broglie wavelength of 775 nm and a detectable photon beam energy below the picowatt level. In comparison, in this paper, even for *N* = 12 and an equivalent de Broglie wavelength of 130 nm, the final detectable energy in the sensing beam is still visible to the naked eye (microwatt level). Therefore, a subtle phase displacement occurs in the insensitive region of the original sensing signal, as shown in Fig. 2b, which can be easily determined using the superresolved signals in real time with only a low-cost detector. Achieving the same performance using the classical interferometric technique would require the use of a deep-ultraviolet laser as the source, which is impossible for most applications.

Another issue of concern is whether and how the OAM carried by state (1), set as $\ell = \pm 1$ for all the results in this paper, affects the measurement performance. First, unlike a superresolution interferometer fed with $N00N$ states, the present scheme does not require the multiphoton amplitude to pass through the phase shifter but provides a highly efficient interface to multiply the phase variation in a classical sensing beam already output from the interferometer. If the phase variation to be measured comes from the interferometric apparatus, such as $\varphi$ loaded by the half-wave plate before the q-plate in Fig. 1a, then the performance of the scheme is independent of the OAM. Second, this conclusion changes if the SOC state itself works as an interferometer and the measured parameter is sensitive to the OAM. For instance, if we replace the birefringence shifter (HW) in Fig. 1a with a Dove prism, then the resolution for measuring the rotation angle of the prism would be proportional to $\ell$. As mentioned previously, the spatial polarization structure of state (1) has an $\ell$-fold angular sensitivity in an image rotation operation [43]. In this type application, using a higher OAM is preferred to pursue extreme performance.

Finally, how can the phase resolution in this scheme be further improved, and what are its limits? According to the above discussion, by adding more upconversion times and using downconversion to push the shortwave limit, we can achieve a higher resolution, i.e., a shorter equivalent de Broglie wavelength $\lambda/N$. However, more parametric interactions mean lower finally detectable signal power. On the one hand, using a high-power laser source combined with a narrow linewidth is better, which is similar to the requirement in classical interferometric apparatuses. On the other hand, optimizing the upconversion strategy to deliver a more favorable OAM mode transformation in parametric processes is crucial. For instance, although we do not have the corresponding crystals for this, cascading another SHG for the $N = 8$ result in Fig. 2b to obtain a super resolution with $N = 16$ is a wiser strategy than the strategy for $N = 12$. This is because the modal weight of $\psi_{\pm N\ell}$ in the final harmonic wave for $N = 16$ is higher than that for $N = 12$ (see *Supplementary Materials*). Notably, using SOC states with type-II SHG can avoid the generation of unwanted mode noise, such as that arising from the transformation from state (1) to state (2), but the latter becomes a pure scalar mode.

This has inspired us to design a device that can convert the latter scalar mode, such as state (2), into an associated SOC mode with its intramodal phase unchanged. In this way, unwanted mode noise can be completely eliminated in the following upconversion, greatly boosting the detectable energy of the final interference signal. In the near future, using this scheme with appropriate technical improvements to achieve a superresolution interferometric measurement with $N > 100$, corresponding to an extreme-ultraviolet de Broglie wavelength, is expected to be an attainable goal.

## Methods

**Laser source.** A narrow linewidth laser operating at 1560 nm (New Focus TLB-6728) was used as the seed light, which can be modulated into 2~10 ns pulses with a 1~10 MHz repetition rate by using a 40 dB electro-optic intensity modulator (EOIM) as needed. Before being output as the laser source for the experiment, the seed was amplified up to 4 watts (average power) by using an erbium-doped fiber amplifier (EDFA). The laser source was collimated into a Gaussian beam with a single-mode fiber, part of which was directly used as the sensing beam, and the rest of which was frequency doubled to 780 nm using a type-0 PPKTP crystal, which was later used for pumping the parametric downconversion. See Supplementary Materials for more details on the experiments.

**Phase information extraction.** The interferometric signal in this scheme is carried by the intramodal phase of spatial modes, that is, the relative phase between conjugate OAM modes. To extract the information of intramodal phase variation, two approaches for characterization of spatial modes were used for the CV beam and OAM harmonic beams. Specifically, we used spatial Stokes tomography to determine the original phase variation $e^{i\Delta\varphi}$ in state (1); for example, the phase distribution shown in the second row of the top inset in Fig. 1a provides the intramodal phase of the SOC state. For scalar harmonic modes obtained in the following upconversion, we used a mode projection measurement based on complex amplitude modulation to extract the superresolved phase variation $e^{iN\Delta\varphi}$ within conjugate OAM modes $\psi_{\pm N\ell}$. Related details are given in the Supplementary Materials.

**Nonlinear crystals.** The nonlinear crystals used in this work include two categories of quasi-phase-matching crystals, i.e., (i) periodically poled KTiOPO$_4$ (PPKTP) crystals with a monoperiodic structure and (ii) periodically poled lithium niobate (PPLN) crystals with a quasiperiodic structure. All the crystals are placed in temperature controllers with a stability of 0.002°C. The periodic parameters of the PPKTP crystals were as follows: (a) 46.1 $\mu m$ cycle designed for type-II SHG $1560\ nm \to 780\ nm$; (b) 7.8 $\mu m$ cycle designed for type-II SHG $780\ nm \to 390\ nm$; (c) 9.1 $\mu m$ cycle designed for type-0 SFG (or difference frequency generation (DFG)) $780\ nm + 1560\ nm \to 520\ nm$; and (d) 2.95 $\mu m$ cycle designed for type-0 SHG $780\ nm \to 390\ nm$. The design parameters of Quasi-periodic MgO-doped PPLN crystals include: two building blocks (A and B) of the quasi-periodic structure $D_A$ and $D_B$, the width of the positive domain in both blocks $l_C$, and $= \tan\vartheta$ with $\vartheta$ being the projection angle which determines the quasi-periodic order. Parameters for two specific Quasi-periodic crystals are (working temperature is 50 ℃): (1) A quasi-periodically poled MgO-doped LiNbO$_3$ to realize a type-II SHG $1560nm(o) + 1560nm(e) \to 780nm(o)$, cascading a type-0 SFG $1560nm(e) + 780nm(e) \to 520nm(e)$, corresponding structure parameters are $D_A = 7.950\mu m$, $D_B = 12.319\mu m$, $l_C = 3.960\mu m$, and $\tau = 2.2212$; the two reciprocal vectors $G_{11}$ and $G_{21}$, are used to compensate for the wave-vector mismatches of the two nonlinear processes, respectively. (2) A quasi-periodically poled MgO-doped LiNbO$_3$ to realize a type-II SHG $1560nm(o) + 1560nm(e) \to 780nm(o)$, cascading a type-0 SHG $780nm(e) + 780nm(e) \to 390nm(e)$, corresponding structure parameters are $D_A = 7.390\mu m$, $D_B = 9.652\mu m$, $l_C = 3.700\mu m$, and $\tau = 0.1801$; the two reciprocal vectors, $G_{11}$ and $G_{43}$, are used to compensate for the wave-vector mismatches of the two nonlinear processes, respectively.

## Acknowledgments

Z.-H. Z., C. R.-G., and B.-S. S. acknowledge support from the National Natural Science Foundation of China (Grant Nos. 62075050, 11934013, and 61975047). X.-P. H. acknowledges support from the National Key R&D Program of China (Nos 2019YFA0705000) and the National Natural Science Foundation of China (Grant Nos. 12174185 and 91950206).